\documentclass[12pt]{article}

\usepackage{scicite}

\usepackage{graphicx}
\usepackage{amsmath}

\usepackage{times}

\newenvironment{sciabstract}{%
\begin{quote} \bf}
{\end{quote}}

\newcounter{lastnote}

\title{Photonic topological phase transition induced by material phase transition}

\author
{Takahiro Uemura,$^{1,2}$ Yuto Moritake,$^{1}$ Taiki Yoda,$^{1,2}$ Hisashi Chiba,$^{1,2}$\\
Yusuke Tanaka,$^{2}$ Masaaki Ono,$^{2,3}$ Eiichi Kuramochi$^{2,3}$ and Masaya Notomi$^{1,2,3\ast}$\\
\\
\normalsize{$^{1}$Department of Physics, Tokyo Institute of Technology, 2-12-1 Ookayama, Meguro,}\\
\normalsize{152-8550, Tokyo, Japan}\\
\normalsize{$^{2}$NTT Basic Research Laboratories, Nippon Telegraph and Telephone Corporation,}\\
\normalsize{3-1 Morinosato-Wakamiya, Atsugi, 243-0198, Kanagawa, Japan}\\
\normalsize{$^{3}$NTT Nanophotonics Center, Nippon Telegraph and Telephone Corporation}\\
\normalsize{3-1 Morinosato-Wakamiya, Atsugi, 243-0198, Kanagawa, Japan}\\
\\
\normalsize{$^\ast$Corresponding author; E-mail:  notomi@phys.titech.ac.jp;}
}

\date{}

\begin{document} 


\baselineskip24pt


\maketitle


\begin{sciabstract}
Photonic topological insulators (PTIs) have been proposed as an analogy to topological insulators in electronic systems. In particular, two-dimensional PTIs have gained attention for the integrated circuit applications. However, controlling the topological phase after fabrication is difficult because the photonic topology requires the built-in specific structures. This study experimentally demonstrates the band inversion in two-dimensional PTI induced by the phase transition of deliberately-designed nanopatterns of a phase-change material, Ge${_2}$Sb${_2}$Te${_5}$ (GST), which indicates the first observation of the photonic topological phase transition with changes in the Chern number. This approach allows us to directly alter the topological invariants, which is achieved by symmetry-breaking perturbation through GST nanopatterns with different symmetry from original PTI. The success of our scheme is attributed to the ultrafine lithographic alignment technologies of GST nanopatterns. These results demonstrate to control photonic topological properties in a reconfigurable manner, providing an insight into new possibilities for reconfigurable photonic processing circuits. 
\end{sciabstract}


\section*{INTRODUCTION}

Different types of solid-state materials possess hidden topological properties in the wavenumber space, which was first inspired by the quantum Hall effects \cite{klitzing1980new} and later led to various novel effects and materials, including topological insulators and edge states \cite{xiao2010berry,hasan2010colloquium,qi2011topological,pesin2012spintronics}. Recently, it has been revealed that periodic dielectric structures also possess analogical topological properties, and the photonic topological insulators (PTIs) have been demonstrated in various systems \cite{wu2015scheme,barik2016two,barik2018topological,khanikaev2017two,hafezi2014topological,joannopoulos1997photonic,wang2009observation,susstrunk2015observation,haldane2008possible,fang2012realizing,khanikaev2013photonic,liang2013optical,rechtsman2013photonic}. Similar to topological electronic systems, topological edge modes are formed in PTIs, which are expected to provide unidirectional propagation with suppressed backscattering and robustness to disorder \cite{wu2015scheme, barik2016two, barik2018topological}. Extensive studies of topological photonics reveal various aspects of light, and topological photonics attracts academic interest as well as  expectations for applications in photonic integrated circuits.

In conventional topological photonics, the topology is generated by the built-in spatial structure; thus, emergent intriguing topological properties are fixed after fabrication. The realization of switchable photonic topological phase will bring new degrees of freedom, such as on-demand edge modes. Therefore, our primary target is to realize controllable photonic topological phase transition in two-dimensional (2D) systems. There have been several theoretical proposals and numerical studies related to reconfigurable topological photonics; however, some of them adopt simple one-dimensional (1D) systems that do not fully utilize the merit of photonic topology \cite{moritake2022, pieczarka2021, takata2018}, and there are various works in 2D systems that just simply shift the frequency of the topological modes by the small refractive index change,  without any access to the topological invariants of the PTIs \cite{cheng2016robust, chen2016, chan2018, Shalaev2018, Dobrykh2018, song2018electrically, cao2019dynamically, Hu2021, Shalaev2019}. Related to our goal, a few numerical studies have been reported for ring-resonator-based PTIs \cite{Leykam2018, kudyshev2019photonic, kudyshev2019tuning} and valley photonic crystal with ferroelectric materials \cite{wu2018reconfigurable} without experimental demonstrations, and these systems face issues, such as large footprint and power consumption.

We propose a method to combine a phase-change material, Ge$_{2}$Sb$_{2}$Te$_{5}$ (GST), with 2D PTIs with $C_6$ rotational symmetry proposed by Wu \textit{et al.} \cite{wu2015scheme, barik2016two, barik2018topological}. $C_6$ PTIs are based on honeycomb-lattice photonic crystals whose lattice constant is approximately half of optical wavelength, significantly smaller than that of ring resonator-based PTIs. 
Because $C_6$ PTIs consist of simple structures with small unit size, it is feasible for integrated-circuit applications. GST exhibits an amorphous-crystal phase transition at approximately 500 $K$, and the refractive index switches between $4.39+0.16i$ for the amorphous (a-) phase and $7.25+1.55i$ for the crystalline (c-) phase at a wavelength of 1.55 {\textmu}m \cite{tanaka2012ultra}. Furthermore, GST can undergo reversible phase transitions induced by by optical pulses with durations of a few tens of nanoseconds \cite{tanaka2012ultra, takeda2014ultrafast}. Therefore, GST is considered a promising material for reconfigurable photonics, and several applications were demonstrated, including optical resonators \cite{rude2013optical}, optical modulators \cite{tanaka2012ultra, zheng2018gst, rios2015integrated}, optical neural network \cite{feldmann2021}, and reconfigurable metasurfaces \cite{chu2016active, nam2007electron}. Despite such advantages of GST and $C_6$ PTI; however, the GST phase transition cannot always be utilized to induce the topological phase transition. For example, it does not work if one simply makes GST-based photonic crystals or loads GST films onto photonic crystals. 

We aim to load a deliberately-designed nanoscale GST thin film pattern on photonic crystals to realize the photonic topological phase transition induced by the material phase transition of GST. Because the nanoscale GST pattern has a different symmetry from that of the host photonic crystal, the refractive index change of GST produces a strong perturbation to the symmetry of the photonic eigenmodes, which induces the photonic topological phase transition. This method requires precise nanoscale patterning of a phase-change material film accurately aligned with the host photonic crystal. Such hybrid nanostructures have not yet been realized; however, in this study, we achieved these hybrid nanostructures using our high-resolution nanofabrication technologies. In this report, we explore the possibility of topological phase switching in PTI with this method, aiming at reconfigurable topological photonic circuits. In addition, the use of material phase transition allows us to exploit non-volatile switching operation, indicating memory functions and energy-saving operations. Furthermore, the proposed tuning method based on nano-patterned phase-change materials is not limited to PTIs; however, it can apply many interesting novel phenomena in nanophotonics originating from the symmetries of nanostructures, such as valley photonics \cite{Dong2017}, bound states in the continuum \cite{Hsu2016}, electromagnetically-induced transparency \cite{RevModPhys.77.633}, and various non-Hermitian optics, including parity-time symmetry \cite{El-Ganainy2018}. Hence, our present method can offer a wide range of applications to control these novel nanophotonic phenomena by the material phase transition. In addition, our achievement provides an intriguing link between photonics and condensed matter physics in a completely novel way, which allows further exploration.

\section*{RESULTS AND DISCUSSION}

\subsection*{Design of GST-loaded topological photonic crystal}

\begin{figure}[ht!]
	\begin{center}
	  \includegraphics[clip,width=16cm]{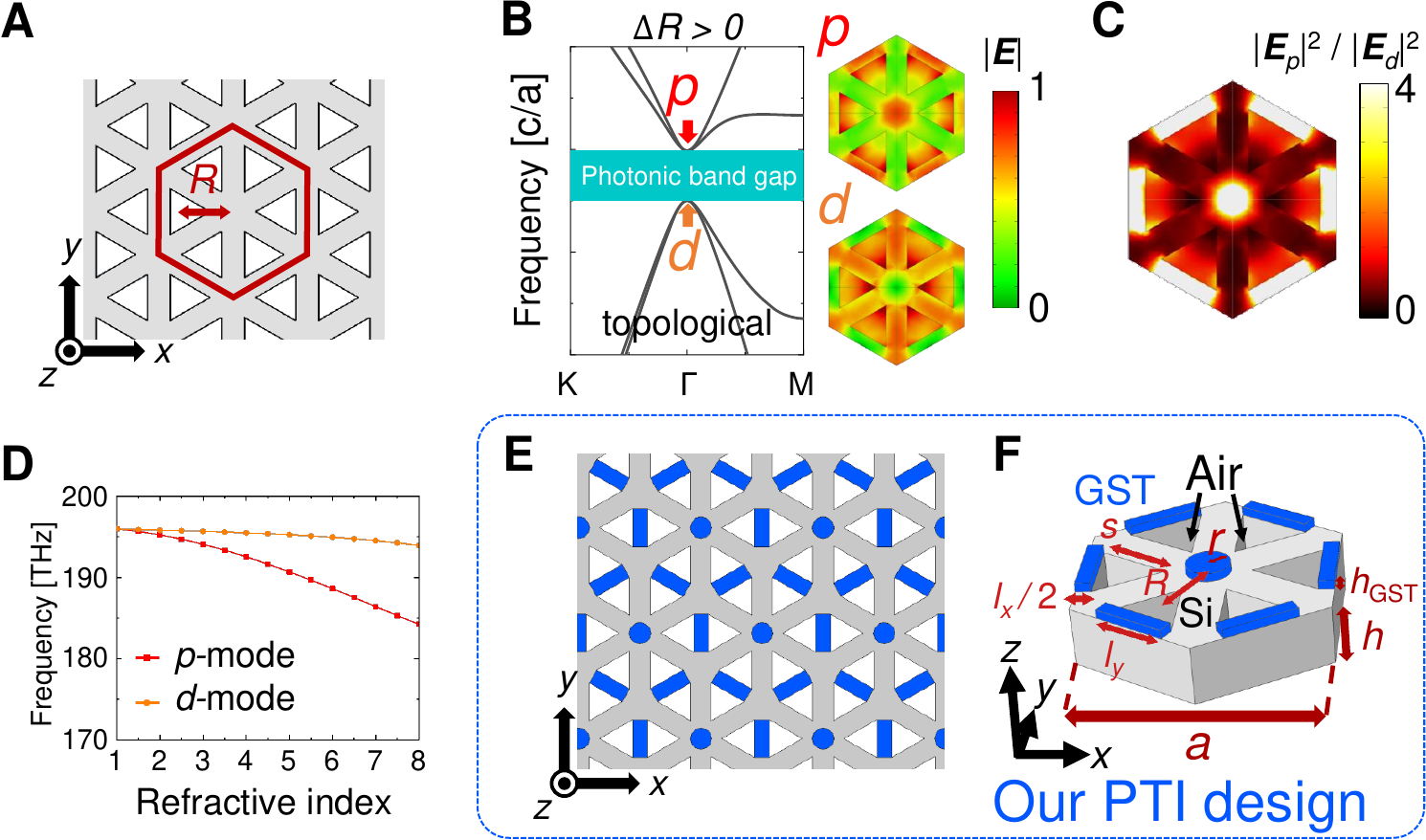}
	  \caption{
		(A) Schematic of a Si topological photonic crystal slab structure. The gray area is Si with $n_{\text{Si}} = 3.48$. The lattice constant $a$ and the slab thickness $h$ are 820 nm and 200 nm, respectively. The side length of a triangle $s$ is $0.31a$. Here, we demonstrate a unit cell with six holes, corresponding to a tripled primitive unit cell of the honeycomb lattice. 
        (B) Dispersion relations of the TE-like mode for the photonic crystal with $\Delta R > 0$, and electric field distribution of the $p$- and $d$- modes at $\Gamma$ point in the unit cell. The plotted electric field distributions represent the sum of two eigenmodes of the effective Hamiltonian at the gamma point, denoted as $\boldsymbol{E}_p := \boldsymbol{E}_{p_x} + \boldsymbol{E}_{p_y}$ for the $p$-mode and $\boldsymbol{E}_d := \boldsymbol{E}_{d_{xy}} + \boldsymbol{E}_{d_{x^2-y^2}}$ for the $d$-mode. The variables $\boldsymbol{E}_{p_x}$, $\boldsymbol{E}_{p_y}$, $\boldsymbol{E}_{d_{xy}}$, and $\boldsymbol{E}_{d_{x^2-y^2}}$ represent the electric field distributions of the basis of the effective Hamiltonian at the $\Gamma$ point. 
        (C) ${\lvert \boldsymbol{E}_p \rvert }^2 / {\lvert \boldsymbol{E}_d \rvert }^2$ distribution in the unit cell. 
        (D) The shift in frequency of $p$-mode and $d$-mode at the $\Gamma$ point as we change the refractive indices of the white region in (C).
        (E) Schematic of the GST-loaded photonic crystal slab. $a = 820$ nm, $s = 0.31a$, $r = 4a/45$, $h = 200$ nm, $h_{\text{GST}} = 30$ nm. The impact of the thickness is described in Supplementary Information S2.2. 
        (F) The unit cell of the photonic crystal slab. 
	  }
	\label{fig1}
	\end{center}
\end{figure}

First, we designed GST-loaded topological photonic crystals, which can exhibit the photonic topological phase transition by the material phase change of GST. All numerical calculations were performed by finite element method (COMSOL Multiphysics). We started with the $C_6$ honeycomb-based PTIs \cite{wu2015scheme,barik2016two,barik2018topological}. Figure \ref{fig1}A shows a photonic crystal slab structure realizing a $C_6$ honeycomb-based PTI. $R$ is the distance from the center of the hexagonal unit cell to the center of a triangular hole. When $R = a/3$, the structure corresponds to a honeycomb lattice, and two pairs of $p$- and $d$-modes become both degenerate and double Dirac cones appear at the $\Gamma$ point. The perturbation of the position of the triangular holes, namely $\Delta R= R - a/3$, breaks the translational symmetry of the honeycomb lattice, thereby tripling the unit cell size, lifting the Dirac degeneracy, and opening a band gap. In particular, this $\Delta R$ perturbation deforms a honeycomb lattice toward a Kagome lattice for $\Delta R > 0$ and a triangular lattice for $\Delta R < 0$ both with a tripled unit cell. For the $\Delta R > 0$ perturbation, namely, when the holes are shifted outwards, so called “expanded” structures, the bands exhibit topologically nontrivial phase owing to the band inversion (Fig. \ref{fig1}B). In the topological phase, the $p$- and $d$-modes are on the upper and lower edge of the topological photonic band gap (PBG). The topological phase transition occurs when the frequency of the $p$-mode ($\omega_p$) and the $d$-modes ($\omega_d$) are reversed; when $\omega_d$ becomes higher than $\omega_p$, the band gap becomes a trivial phase. The topological phase of the system is defined by the Chern number of the eigenstates, which assumes a value of $0$ in the trivial phase and $\pm 1$ in the nontrivial phase (See Supplementary Information S1 for details).

Our key approach is to deliberately place the phase change material on the photonic crystal in such a way that the material phase change induces the symmetry breaking perturbation in an effectively similar way to the symmetry breaking by the hole shift $\Delta R$. The frequency shift owing to the refractive index change depends on the magnitude of the spatial overlap between the electromagnetic field of the mode and the material. Figure \ref{fig1}C shows the ratio of the electric field intensity distribution of the two $p$- and $d$-modes. In Fig. \ref{fig1}C, the white area corresponds to the largest field intensity contrast between the $p$- and $d$-modes, that is, the region where the $p$-modes dominate over the $d$-modes. Therefore, if a phase change material is placed in that area, $\omega_p$ should be dominantly shifted with a negligible shift of $\omega_d$. Figure \ref{fig1}D shows the frequency change of each mode when the refractive index in the white area on the unit cell shown in Fig. \ref{fig1}C is artificially changed, and apparently the large shift is obtained only for $\omega_p$. Note that the geometric size of the blue area is considerably smaller than the period of the photonic crystal. Therefore, sub-wavelength patterning of GST is essentially needed to produce the symmetry-breaking perturbation. Note that the blue area causes a perturbation that preserves the $C_6$ symmetry but breaks the translational symmetry of the $\Delta R = 0$ honeycomb lattice. A crucial aspect of our proposed scheme is that both the hole shift $\Delta R > 0$ and the phase transition of nano-patterned GST to the crystalline phase ($\Delta n > 0$) result in the same symmetry being broken in the reverse direction. Consequently, incorporating $\Delta R$ as a pre-tuning in the structure allows for the band inversion of the frequencies of the $p$- and $d$-modes, which is caused by the perturbation effect of the phase change of GST. 
This scheme allows for the direct access to the Chern number, i.e., photonic topological phase transitions, associated with the refractive index change of nanoscale GST (See Supplementary Information S1). We highlight the distinction from previous devices that employ uniform refractive index modulation to alter the frequency without accessing the Chern number\cite{cheng2016robust, chen2016, chan2018, Shalaev2018, Dobrykh2018, song2018electrically, cao2019dynamically, Hu2021, Shalaev2019}.

\begin{figure}[ht!]
	\begin{center}
	  \includegraphics[clip,width=16cm]{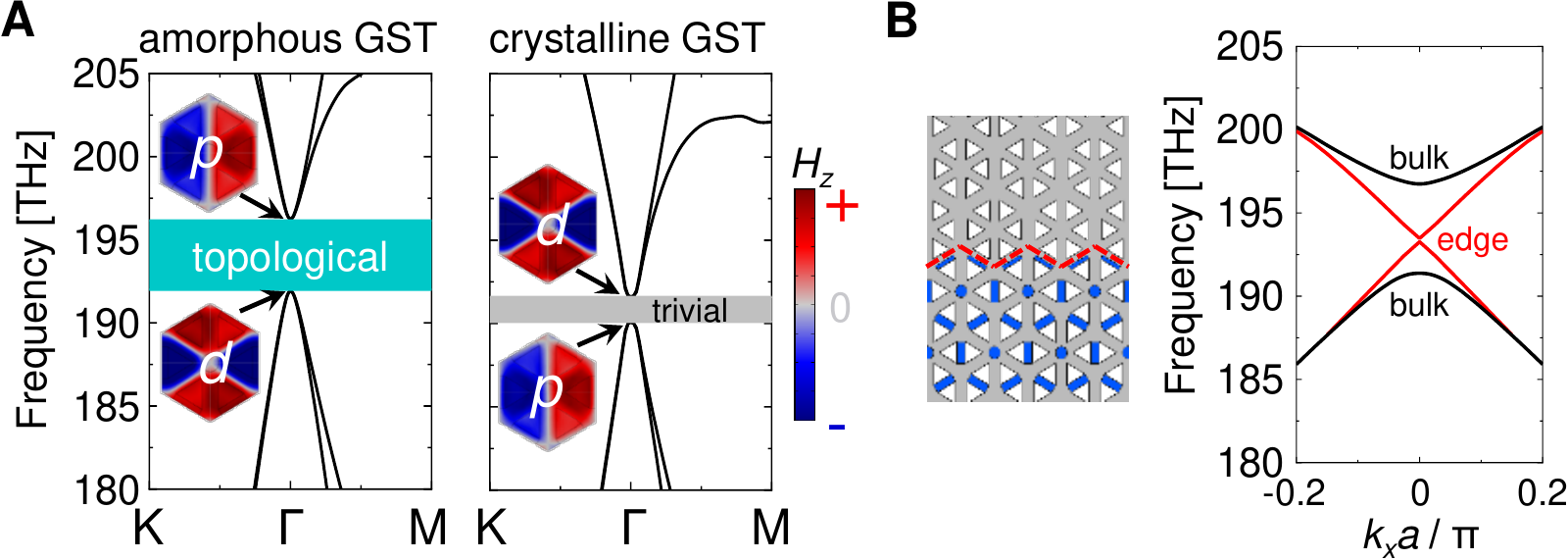}
	  \caption{
        (A) Band diagrams of the TE-like mode for $n_{\text{amo}} = 4.39 + 0.16i$ (left) and $n_{\text{cry}} = 7.25 + 1.55i$ (right) and corresponding $H_z$ distributions at the $\Gamma$ point. Here, “$p$” and “$d$” show the $H_z$ distributions of $p_x$ and $d_{x^2-y^2}$, respectively.
        (B) Band diagrams of the TE-like mode in the heterostructures for the amorphous phases. The red lines indicate the topological edge modes. 
	  }
	\label{fig1a}
	\end{center}
\end{figure}

Figures \ref{fig1}E and \ref{fig1}F show the structure of our proposed GST-loaded topological photonic crystal. The host silicon photonic crystal is designed to be in a topological phase ($\Delta R = (16/1000)a$), to ensure the phase is switched to a trivial phase by material phase transition. Figure \ref{fig1a}A shows the photonic band diagrams and calculated $H_z$ distributions at the $\Gamma$ point in the amorphous phase (a-phase) and the crystalline phase (c-phase). The $H_z$ distribution at the $\Gamma$ point shows that we can achieve $\omega_p > \omega_d$ in the a-phase, while $\omega_p < \omega_d$ in the c-phase. Because the phase of a PBG is determined by $\omega_p$ and $\omega_d$ at the $\Gamma$ point, the topology of the PBG is non-trivial at the a-phase and trivial at the c-phase. Therefore, this numerical simulation shows that the present structure achieves topological phase transition by material phase transition if the designed sub-wavelength GST pattern is appropriately placed. See Supplementary Information S2.3 for the impact of the imaginary part of the refractive index of GST. Generally, topological edge (domain wall) modes appear at the boundary between crystals with different topologies. The simulated photonic band of the heterostructure shown in Fig. \ref{fig1a}B demonstrates that edge modes occur only when a topological gap exists (See Supplementary Information S2.4 for details).

\subsection*{Experimental demonstration of photonic topological phase transition}

\begin{figure}[ht!]
	\begin{center}
	  \includegraphics[clip,width=12cm]{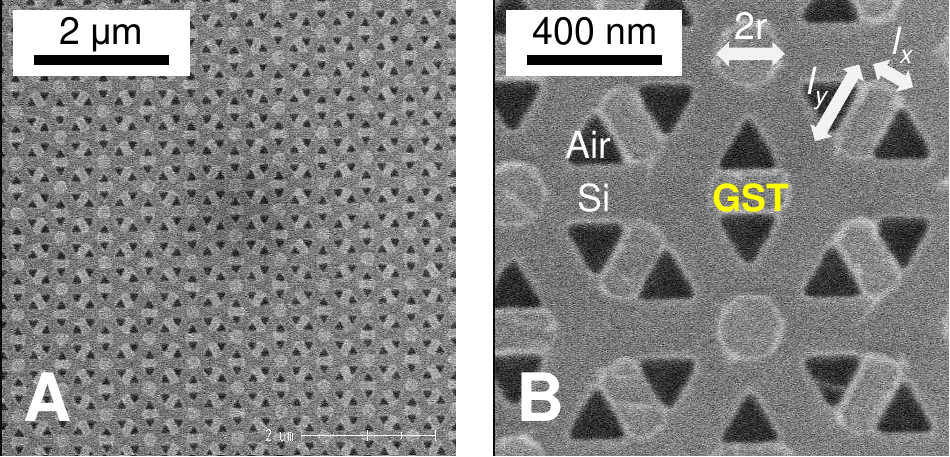}
	  \caption{
		(A), (B) SEM images of the fabricated photonic crystal with the Ge$_{2}$Sb$_{2}$Te$_{5}$ film. The radius of the GST circles $r$ is approximately 90 nm, and the length of the GST rectangles $l_x$, $l_y$ are approximately 120 nm and 200 nm, respectively.
        }
	  \label{fig2}
	\end{center}
\end{figure}
Here, we experimentally demonstrate the topological phase transition in fabricated photonic crystals. The details of the fabrication and measurement are described in the Methods section. The designed Si photonic crystal structure was fabricated by e-beam lithography and dry etching. The patterned GST film was precisely formed on the Si photonic crystals by additional e-beam lithography using the lift-off process. The GST pattern sizes, $r$ and $s$, were approximately 90 and 200 nm, respectively. The most critical aspect of the fabrication is alignment. We achieved alignment accuracy better than 10 nm. The scanning electron microscopic (SEM) images (Figs. \ref{fig2}A and \ref{fig2}B) showed that good alignment accuracy could be achieved. To the best of our knowledge, the patterning of GST on nanostructured materials at this size scale has not been demonstrated.

\begin{figure}[ht!]
	\begin{center}
	  \includegraphics[clip,width=15cm]{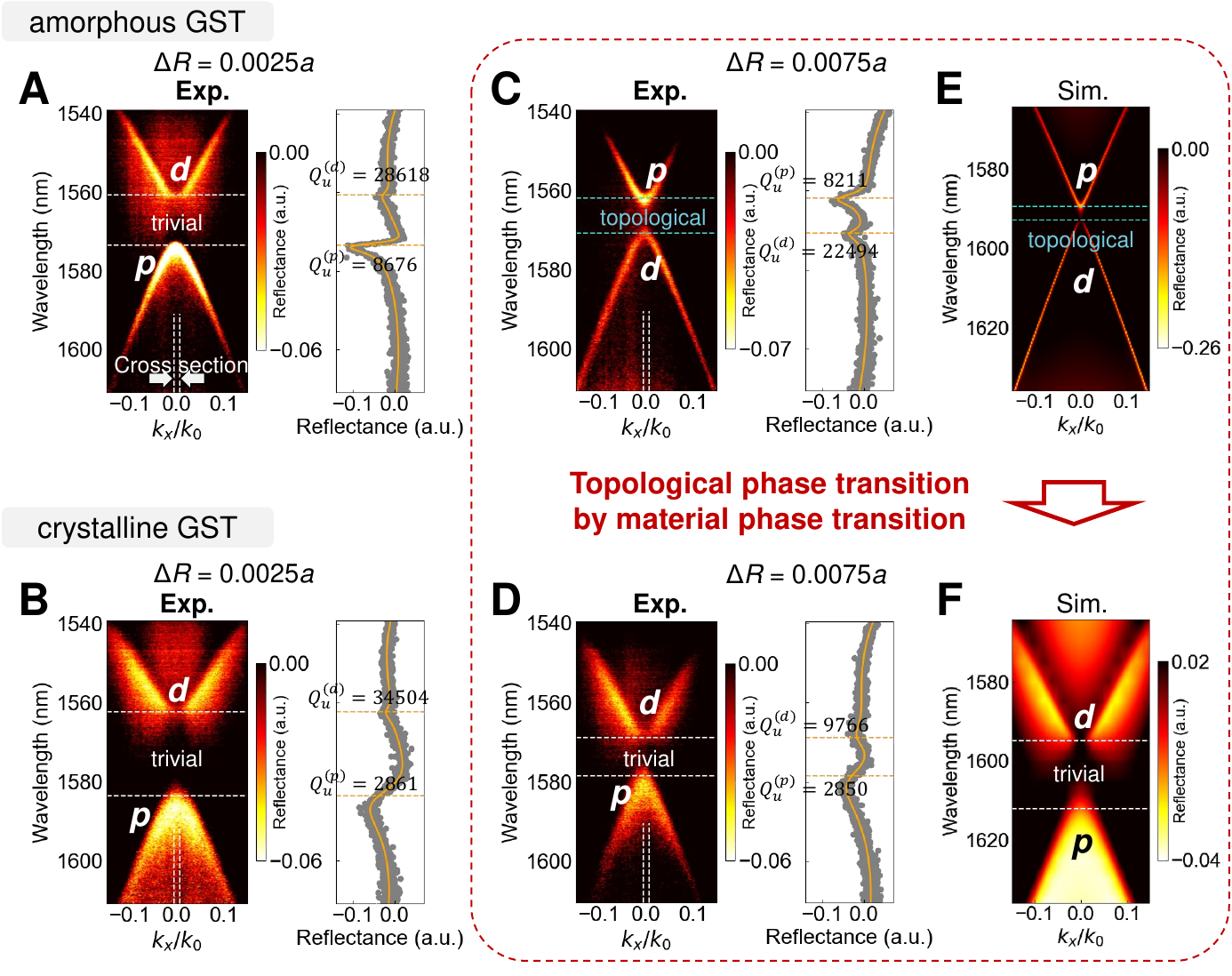}
	  \caption{
		(A-D) Measured reflection spectra along the $k_x$ direction (A) before (GST is in amorphous phase) and (B) after (crystalline phase) annealing for the sample of $\Delta R=0.0025a$, and (C), (D) show the same as (A), (B) for the sample of $\Delta R=0.0075a$. The right panels show the cross section of the measured band diagram at the $\Gamma$ point. We fitted these curves using temporal coupled mode theory (TCMT), and deduce the out-of-plane radiation quality factor of the band ($Q_u$). (E), (F) Simulated reflection spectra along the $k_x$ direction in (E) amorphous and (F) crystalline GST. 
		  }
	  \label{fig3}
	\end{center}
\end{figure}
Photonic band diagrams of the fabricated structures were measured by angle-resolved reflection spectroscopy \cite{ZHANG2021824}, as shown in the left panels of Figs. \ref{fig3}A to \ref{fig3}D. Two bands together with a band gap were visible. To determine whether the gap is topological or not, we noted the brightness of the band at the $\Gamma$ point. Owing to the difference in symmetry, the $p$-mode should have more out-of-plane radiation, meaning brighter than the $d$-mode to detect the band inversion and distinguish the topological phase. As a first example, see Fig. \ref{fig3}A for $\Delta R=0.0025a$ in the a-phase (as-deposited), where the lower band is apparently brighter. Thus, the lower band is the $p$-mode and the upper is the $d$-mode. Figure \ref{fig3}D shows a band diagram for $\Delta R=0.0025a$ after annealing above the phase-change temperature of GST; thus, in the c-phase. We can observe the same band configuration, indicating that $\Delta R=0.0025a$ remains trivial irrespective of the GST phase.

The situation significantly changes for $\Delta R=0.0075a$, as shown in Fig. \ref{fig3}C in the a-phase and Fig. \ref{fig3}D in the c-phase. In Fig. \ref{fig3}C, the upper band exhibits a greater degree of reflection intensity than that of the lower mode at the $\Gamma$ point, which reflects the contrast in out-of-plane radiation loss of the eigenmodes. Conversely, in Fig. \ref{fig3}D, the lower band exhibits a greater degree of reflection intensity at the $\Gamma$ point. Given that the $p$-mode has a greater out-of-plane radiation loss, this result strongly indicates that $\omega_d < \omega_p$ in the a-phase and $\omega_d > \omega_p$ in the c-phase. This demonstrates that the band inversion takes place owing to the phase transition of GST, and implies that the band gap is trivial in the c-phase and topological in the a-phase, which is the aim of this study. The measured results are consistent with the numerical simulations as shown in Figs. \ref{fig3}E and \ref{fig3}F. Furthermore, we quantitatively investigated this phenomenon. The right panels in Figs. \ref{fig3}A to \ref{fig3}D show the cross section of the measured band diagram at the $\Gamma$ point, and the estimated out-of-plane radiation Q factor ($Q_u$) are shown in the panels. We fitted these curves using temporal coupled mode theory (TCMT) and extracted the upper out-of-plane radiation Q factor ($Q_u$) (See Supplementary Information S5 and Fig. S11 for details). We can observe a $Q_u$ contrast between two bands in Figs. \ref{fig3}A to \ref{fig3}D, which confirms our band assignment of $p$- and $d$-modes. The upper out-of-plane radiation Q factor is apparently inverted for $\Delta R=0.0075a$ between the c- and a-phases in Figs. \ref{fig3}B and \ref{fig3}D, demonstrating the band inversion. These results verify that the material phase transition of GST between the amorphous and crystalline phases has induced the photonic topological phase transition.

\begin{figure}[ht!]
	\begin{center}
	  \includegraphics[clip,width=15cm]{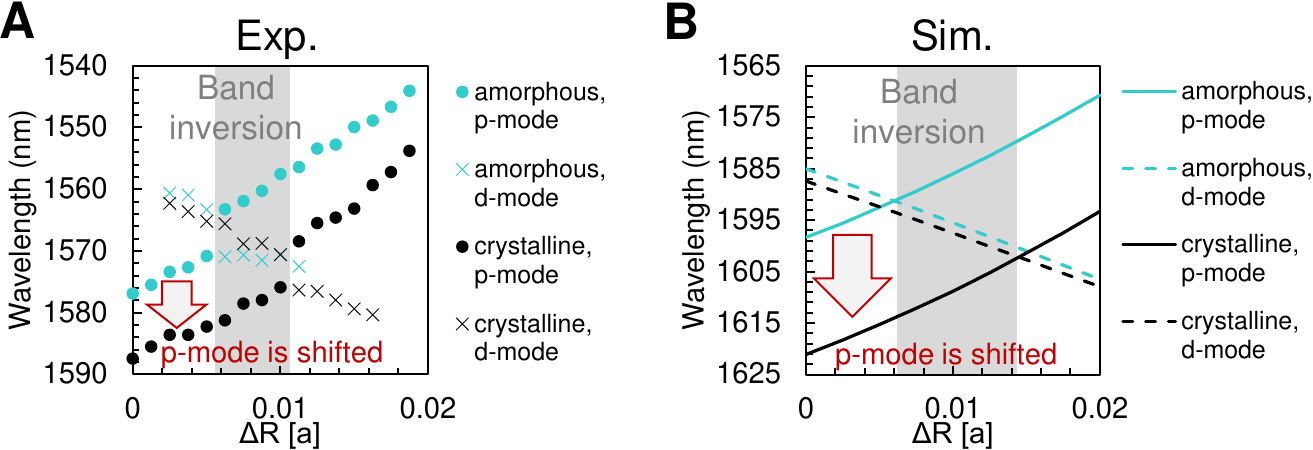}
	  \caption{
		(A), (B) Frequency (wavelength) of the photonic band at the $\Gamma$ point as a function of the hole shift $\Delta R$ in (A) experiments and (B) corresponding simulations.
		  }
	  \label{fig3a}
	\end{center}
\end{figure}

We analyze the measured band diagrams for various $\Delta R$ values, and assign each mode as $p$- or $d$-mode based on the deduced $Q_u$ contrast. Figures \ref{fig3a}A and \ref{fig3a}B show the experimental and simulated wavelength shifts of the $p$- and $d$-modes at the $\Gamma$ point as a function of $\Delta R$. As shown in Fig. \ref{fig3a}A, only the $p$-mode is selectively shifted by the phase change of GST, which is attributed to our GST pattern design shown in Fig. \ref{fig1}E. For certain values of $\Delta R$, the out-of-plane radiation Q factor of the $d$-mode was insufficient to permit the acquisition of spectra through fitting, and thus the corresponding points are not displayed in Fig. \ref{fig3a}A (See Supplementary Information S5 and Fig. S10). The shaded area indicates the parameter range where the photonic topological phase is changed by the material phase transition. We can observe that the achieved topological phase transition relies on appropriate $\Delta R$, which supports our design methodology. In Fig. \ref{fig3a}A, we can notice a slight anti-crossing behavior of two bands near the boundary of parameters. This indicates a small coupling of $p$- and $d$-mode, which might be caused by the fabrication disorder of GST blocks. However, an overall good agreement between experiments (Fig. \ref{fig3a}A) and simulations (Fig. \ref{fig3a}B) is obtained, indicating that our experimental demonstration can be explained by the proposed mechanism of topological phase transition via material phase transition.

\begin{figure}[ht!]
	\begin{center}
	  \includegraphics[clip,width=16cm]{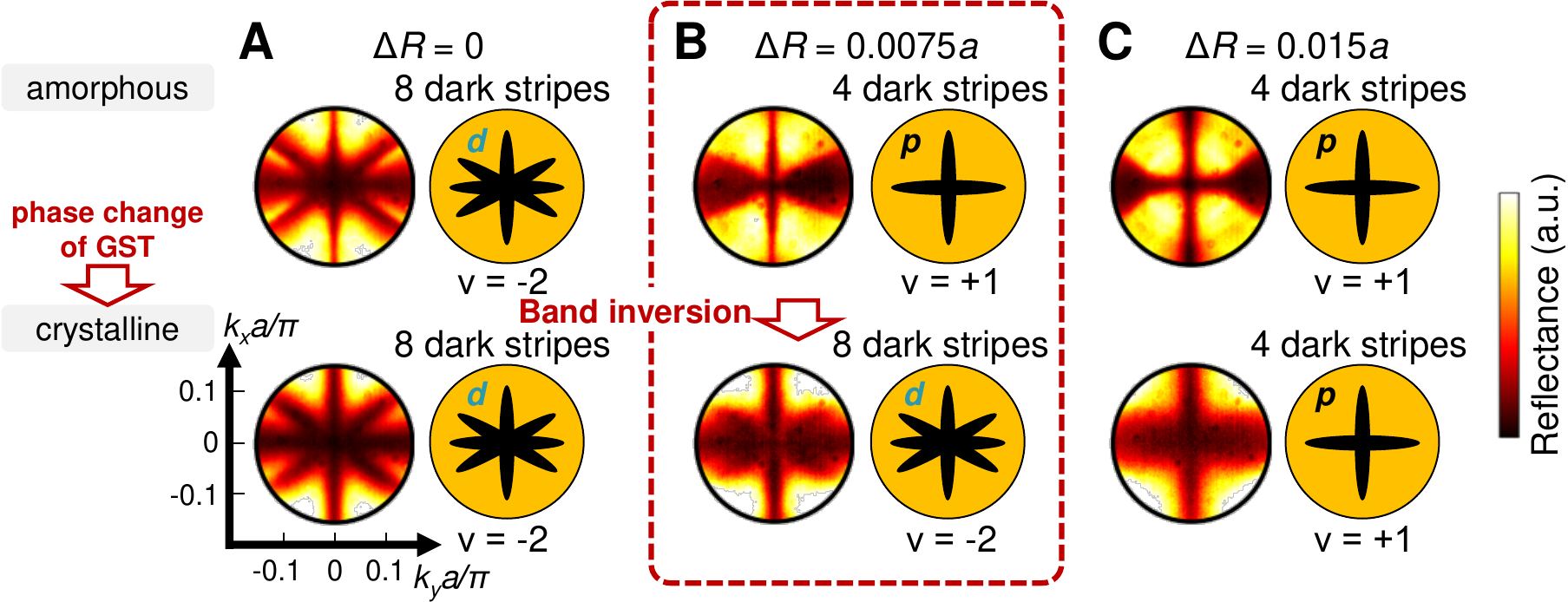}
	  \caption{
        Measured Fourier image for the sample of (A) $\Delta R=0$, (B)     $\Delta R=0.0075a$, (C) $\Delta R=0.015a$ when the two polarizers are perpendicular. The top and bottom rows represent the results in the amorphous and crystalline phases, respectively.
		  }
	  \label{fig4}
	\end{center}
\end{figure}
Finally, we examine the polarization properties at the material phase transition. It has been established that the $p$- and $d$-modes in this photonic crystal have distinct topological charges for polarization of $+1$ and $-2$ around the $\Gamma$ point, because the honeycomb lattice has $C_6$ symmetry \cite{ji2022probing}. Thus, band inversion should induce changes in the polarization state of the bands. Here, we intend to clarify the topological properties of our sample from the polarization measurement. To investigate the polarization state changes associated with the GST phase transition, the Fourier images of the photonic bands were measured using a pair of polarizers in the crossed-nicols setup. Figures \ref{fig4}A to \ref{fig4}C show the measured Fourier plane images on the upper bands. To better visualize the dark stripes, we integrated multiple Fourier plane images measured with bandpass filters for multiple wavelength ranges (See Supplementary Information S6.2 for details). Note that this crossed-nicols polarization measurement delivers the information about the polarization symmetry, which is directly related to the topological charge. The symmetry argument states that when a band at the $\Gamma$ point is in $p$-mode ($d$-mode), four (eight) dark stripes should appear in the Fourier image at the crossed-nicols setup. In the sample with $\Delta R=0$ shown in Fig. \ref{fig4}A, eight dark stripes appear in both the a- and c-phases, indicating that the topological charge of this upper band around the $\Gamma$ point is $-2$. This result reveals that the upper band is the $d$-mode (meaning a topologically trivial gap) regardless of the phase of the GST, and is consistent with the results in Fig. \ref{fig3a}A. However, four dark stripes are observed in the sample with $\Delta R=0.015a$. Therefore, the topological charge of the upper band is $+1$, that is, this sample has a topologically non-trivial bandgap. In the results of the sample with $\Delta R=0.0075a$, a change in the number of nodes was observed in the course of the GST phase change: There are four nodes in the a-phase, indicating that it is the $p$-mode, while eight nodes appear in the c-phase, indicating that it is the $d$-mode. Hence, this result is another proof of the band inversion between the $p$- and $d$-modes via the material phase transition.

\subsection*{Conclusion}

In this report, we proposed and demonstrated, both theoretically and experimentally, a method to achieve the photonic topological phase transition induced by the material phase transition in 2D PTI systems. In particular, we demonstrated that the topological phase transition can occur if we selectively load a thin film of GST on a specific region of a Si PTI. With a precise alignment of nanoscale patterned GST on the photonic crystal, we successfully fabricated and experimentally observed the switching between the topologically non-trivial and trivial phases induced by the structural phase change of GST. GST is a well-established phase change material in which each phase is stable for a long time and the phase transition can be externally controlled in a reversible way by optical pulse excitation at room temperature. Recently, pulse-driven reversible phase change has been employed in many photonic integration applications \cite{tanaka2012ultra, takeda2014ultrafast, rude2013optical, zheng2018gst, rios2015integrated, feldmann2021, chu2016active, nam2007electron}. Therefore, when combined with pulse-driven phase change of GST, the photonic topological phase can be reconfigurably changed after fabrication by the proposed GST-loaded photonic crystals. This potential may open up a wide range of possibilities for applications. For example, the present method may lead to reconfigurable photonic integrated circuits. As we showed, one can generate and eliminate edge waveguiding modes on-demand by manipulating the Chern number through the material phase transition, which we have numerically confirmed in this report. Since we employ the material phase transition, the switched state can be sustained without any power input. This means that each reconfigured circuit can be sustained without extra power. This non-volatile operation not only saves energy consumption but also indicates memory functionalities. Among various forms of PTIs, $C_6$ honeycomb-based PTIs have the small unit size and are compatible to conventional photonic integrated circuits. Consequently, we believe that the proposed GST-loaded PTI is especially suited for applications to low-power reconfigurable photonic integrated circuits. In addition, the present method itself can be applied to other photonic topological systems, such as valley photonics, in which the topological properties are determined by the fabricated structures. We believe that the demonstrated controllability of photonic topology paves the way for novel reconfigurable topological photonic circuits, where reconfigurable topological properties are used as new degrees of freedom.

\section*{MATERIALS AND METHODS}

\subsection*{Fabrication}

To demonstrate the band inversion in the frequency position of $p$-mode and $d$-mode, we fabricated honeycomb lattice photonic crystals and deposited GST films. We experimentally observed a band dispersion by angular-resolved scattering method. GST-loaded photonic crystal slabs were fabricated from a Silicon-On-Insulator (SOI) wafer with a 205 nm Si layer. The resist patterns for the photonic crystal layer were defined by an electron-beam (EB) lithography system and formed by a dry etching process. The resist patterns for the GST films were also defined by an EB lithography system. Approximately 30-nm GST films were deposited by magnetron sputtering using an alloy target in an Ar atmosphere of $1.5 \times 10^{-2}$ Pa and a rate of $0.19$ nm/s on photonic crystal patterns. The deposition rate was measured by the quartz resonator. A patterned GST film was formed by a lift-off process. After some measurements in the as-deposited (amorphous) state, the resonator samples were annealed to crystallize GST by thermal annealing on a hot plate at a temperature of $200^\circ$C, exceeding the crystallization temperature of $\sim 170^\circ$C \cite{jost2015disorder, sumikura2019highly}, and with a duration of 30 minutes. We set the side length of a triangle $s$ to $0.24a$, which is smaller than those in the numerical calculations shown in previous chapters to enlarge the area of the GST blocks fully. We changed the lattice constant to $a = 750$ nm to adjust the wavelength of the photonic band to the telecommunication wavelength. The hole shift $R$ was changed to $\Delta R = 0 \sim 1.875 \ [a/100]$. Fig. \ref{fig3}B shows the scanning electron microscope (SEM) image of the honeycomb lattice photonic crystal after the lift-off process. We confirmed that the GST is correctly patterned in the center of the unit cell and between the triangular hole.

\subsection*{Experimental setup}

Figure S9A in the Supplementary Information S5 shows the schematic of the experimental setup for the band measurement. The unpolarized incident light was illuminated by a tungsten-halogen lamp (Thorlabs, SLS302). The incident light was focused on the sample by an aspherical lens (Thorlabs, C105TMD-C, NA=0.6). The back focal plane of the lens was imaged onto the slit of an imaging spectrometer (Teledyne Princeton Instruments, IsoPlane 320) with an InGaAs camera (Teledyne Princeton Instruments, NIRvana HS) using two convex lenses L1 and L2. The focal lengths of L1 and L2 were 20 cm and 30 cm, respectively.
Fig. S12A in the Supplementary Information S6.2 shows the schematic of the experimental setup for Fourier plane measurement. The basic configuration was identical to the band measurements shown in Fig. 5 of the main text. The polarization of the incident light was chosen by the first polarizer (POL1), and the second polarizer (POL2) was positioned orthogonal to POL1. Bandpass filters (Thorlabs FBH series, FWHM bandwidth: 12 nm) were used to measure Fourier planes for a specific frequency. An InGaAs camera (Teledyne Princeton Instruments, NIRvana 640) was used for the Fourier plane measurement.

\section*{SUPPLEMENTARY MATERIALS}
Supplementary material for this article is available.
Supplemental refers \cite{wu2015scheme, barik2016two, barik2018topological, zhang2018broadband, zhang2019broadband,jiang2018nonvolatile, tanaka2012ultra, wu2018reconfigurable, Hsu2013, PhysRevLett.113.257401, PhysRevResearch.3.L022013, PhysRevLett.131.053802, PhysRevB.108.085116, doi:10.1126/sciadv.aaw4137}.

\section*{Acknowledgments}

We acknowledge invaluable contribution from Dr. Toshiaki Tamamura and Toshifumi Watanabe for nanofabrication processes and fruitful discussions with Dr. Kenta Takata. 
The part of this work is supported by KAKENHI JP20H05641 and JP21K14551.



\bibliography{scibib}

\begin{thebibliography}{10}

\bibitem{klitzing1980new}
K.~v. Klitzing, G.~Dorda, M.~Pepper, {\it Phys. Rev. Lett.\/} {\bf 45}, 494 (1980).

\bibitem{xiao2010berry}
D.~Xiao, M.-C. Chang, Q.~Niu, {\it Rev. Mod. Phys.\/} {\bf 82}, 1959 (2010).

\bibitem{hasan2010colloquium}
M.~Z. Hasan, C.~L. Kane, {\it Rev. Mod. Phys.\/} {\bf 82}, 3045 (2010).

\bibitem{qi2011topological}
X.-L. Qi, S.-C. Zhang, {\it Rev. Mod. Phys.\/} {\bf 83}, 1057 (2011).

\bibitem{pesin2012spintronics}
D.~Pesin, A.~H. MacDonald, {\it Nature Materials\/} {\bf 11}, 409 (2012).

\bibitem{wu2015scheme}
L.-H. Wu, X.~Hu, {\it Phys. Rev. Lett.\/} {\bf 114}, 223901 (2015).

\bibitem{barik2016two}
S.~Barik, H.~Miyake, W.~DeGottardi, E.~Waks, M.~Hafezi, {\it New Journal of Physics\/} {\bf 18}, 113013 (2016).

\bibitem{barik2018topological}
S.~Barik, {\it et~al.\/}, {\it Science\/} {\bf 359}, 666 (2018).

\bibitem{khanikaev2017two}
A.~B. Khanikaev, G.~Shvets, {\it Nature Photonics\/} {\bf 11}, 763 (2017).

\bibitem{hafezi2014topological}
M.~Hafezi, J.~M. Taylor, {\it Physics Today\/} {\bf 67}, 68 (2014).

\bibitem{joannopoulos1997photonic}
J.~Joannopoulos, P.~R. Villeneuve, S.~Fan, {\it Solid State Communications\/} {\bf 102}, 165 (1997). Highlights in Condensed Matter Physics and Materials Science.

\bibitem{wang2009observation}
Z.~Wang, Y.~Chong, J.~D. Joannopoulos, M.~Solja{\v{c}}i{\'{c}}, {\it Nature\/} {\bf 461}, 772 (2009).

\bibitem{susstrunk2015observation}
R.~Süsstrunk, S.~D. Huber, {\it Science\/} {\bf 349}, 47 (2015).

\bibitem{haldane2008possible}
F.~D.~M. Haldane, S.~Raghu, {\it Phys. Rev. Lett.\/} {\bf 100}, 013904 (2008).

\bibitem{fang2012realizing}
K.~Fang, Z.~Yu, S.~Fan, {\it Nature Photonics\/} {\bf 6}, 782 (2012).

\bibitem{khanikaev2013photonic}
A.~B. Khanikaev, {\it et~al.\/}, {\it Nature Materials\/} {\bf 12}, 233 (2013).

\bibitem{liang2013optical}
G.~Q. Liang, Y.~D. Chong, {\it Phys. Rev. Lett.\/} {\bf 110}, 203904 (2013).

\bibitem{rechtsman2013photonic}
M.~C. Rechtsman, {\it et~al.\/}, {\it Nature\/} {\bf 496}, 196 (2013).

\bibitem{moritake2022}
Y.~Moritake, M.~Ono, M.~Notomi, {\it Nanophotonics\/}  (2022).

\bibitem{pieczarka2021}
M.~Pieczarka, {\it et~al.\/}, {\it Optica\/} {\bf 8}, 1084 (2021).

\bibitem{takata2018}
K.~Takata, M.~Notomi, {\it Phys. Rev. Lett.\/} {\bf 121}, 213902 (2018).

\bibitem{cheng2016robust}
X.~Cheng, {\it et~al.\/}, {\it Nature Materials\/} {\bf 15}, 542 (2016).

\bibitem{chen2016}
Z.-G. Chen, Y.~Wu, {\it Phys. Rev. Applied\/} {\bf 5}, 054021 (2016).

\bibitem{chan2018}
H.-C. Chan, G.-Y. Guo, {\it Phys. Rev. B\/} {\bf 97}, 045422 (2018).

\bibitem{Shalaev2018}
M.~I. Shalaev, S.~Desnavi, W.~Walasik, N.~M. Litchinitser, {\it New Journal of Physics\/} {\bf 20}, 023040 (2018).

\bibitem{Dobrykh2018}
D.~A. Dobrykh, A.~V. Yulin, A.~P. Slobozhanyuk, A.~N. Poddubny, Y.~S. Kivshar, {\it Phys. Rev. Lett.\/} {\bf 121}, 163901 (2018).

\bibitem{song2018electrically}
Z.~Song, H.~Liu, N.~Huang, Z.~Wang, {\it Journal of Physics D: Applied Physics\/} {\bf 51}, 095108 (2018).

\bibitem{cao2019dynamically}
T.~Cao, {\it et~al.\/}, {\it Science Bulletin\/} {\bf 64}, 814 (2019). SPECIAL TOPIC: Electromagnetic Metasurfaces: from Concept to Applications.

\bibitem{Hu2021}
W.~Hu, J.~Hu, S.~Wen, Y.~Xiang, {\it Opt. Lett.\/} {\bf 46}, 2589 (2021).

\bibitem{Shalaev2019}
M.~I. Shalaev, W.~Walasik, N.~M. Litchinitser, {\it Optica\/} {\bf 6}, 839 (2019).

\bibitem{Leykam2018}
D.~Leykam, S.~Mittal, M.~Hafezi, Y.~D. Chong, {\it Phys. Rev. Lett.\/} {\bf 121}, 023901 (2018).

\bibitem{kudyshev2019photonic}
Z.~A. Kudyshev, A.~V. Kildishev, A.~Boltasseva, V.~M. Shalaev, {\it Nanophotonics\/} {\bf 8}, 1349 (2019).

\bibitem{kudyshev2019tuning}
Z.~A. Kudyshev, A.~V. Kildishev, A.~Boltasseva, V.~M. Shalaev, {\it ACS Photonics\/} {\bf 6}, 1922 (2019).

\bibitem{wu2018reconfigurable}
Y.~Wu, X.~Hu, Q.~Gong, {\it Phys. Rev. Mater.\/} {\bf 2}, 122201 (2018).

\bibitem{tanaka2012ultra}
D.~Tanaka, {\it et~al.\/}, {\it Opt. Express\/} {\bf 20}, 10283 (2012).

\bibitem{takeda2014ultrafast}
J.~Takeda, W.~Oba, Y.~Minami, T.~Saiki, I.~Katayama, {\it Applied Physics Letters\/} {\bf 104}, 261903 (2014).

\bibitem{rude2013optical}
M.~Rudé, {\it et~al.\/}, {\it Applied Physics Letters\/} {\bf 103}, 141119 (2013).

\bibitem{zheng2018gst}
J.~Zheng, {\it et~al.\/}, {\it Opt. Mater. Express\/} {\bf 8}, 1551 (2018).

\bibitem{rios2015integrated}
C.~R{\'i}os, {\it et~al.\/}, {\it Nature Photonics\/} {\bf 9}, 725 (2015).

\bibitem{feldmann2021}
J.~Feldmann, {\it et~al.\/}, {\it Nature\/} {\bf 589}, 52 (2021).

\bibitem{chu2016active}
C.~H. Chu, {\it et~al.\/}, {\it Laser \& Photonics Reviews\/} {\bf 10}, 986 (2016).

\bibitem{nam2007electron}
S.-W. Nam, {\it et~al.\/}, {\it Journal of The Electrochemical Society\/} {\bf 154}, H844 (2007).

\bibitem{Dong2017}
J.-W. Dong, X.-D. Chen, H.~Zhu, Y.~Wang, X.~Zhang, {\it Nature Materials\/} {\bf 16}, 298 (2017).

\bibitem{Hsu2016}
C.~W. Hsu, B.~Zhen, A.~D. Stone, J.~D. Joannopoulos, M.~Solja{\v{c}}i{\'{c}}, {\it Nature Reviews Materials\/} {\bf 1}, 16048 (2016).

\bibitem{RevModPhys.77.633}
M.~Fleischhauer, A.~Imamoglu, J.~P. Marangos, {\it Rev. Mod. Phys.\/} {\bf 77}, 633 (2005).

\bibitem{El-Ganainy2018}
R.~El-Ganainy, {\it et~al.\/}, {\it Nature Physics\/} {\bf 14}, 11 (2018).

\bibitem{ZHANG2021824}
Y.~Zhang, {\it et~al.\/}, {\it Science Bulletin\/} {\bf 66}, 824 (2021).

\bibitem{ji2022probing}
C.-Y. {Ji}, {\it et~al.\/}, {\it arXiv e-prints\/} p. arXiv:2210.17081 (2022).

\bibitem{jost2015disorder}
P.~Jost, {\it et~al.\/}, {\it Advanced Functional Materials\/} {\bf 25}, 6399 (2015).

\bibitem{sumikura2019highly}
H.~Sumikura, {\it et~al.\/}, {\it Nano Letters\/} {\bf 19}, 2549 (2019).

\bibitem{zhang2018broadband}
Q.~Zhang, {\it et~al.\/}, {\it Opt. Lett.\/} {\bf 43}, 94 (2018).

\bibitem{zhang2019broadband}
Y.~Zhang, {\it et~al.\/}, {\it Nature Communications\/} {\bf 10}, 4279 (2019).

\bibitem{jiang2018nonvolatile}
W.~Jiang, {\it Scientific Reports\/} {\bf 8}, 15946 (2018).

\bibitem{Hsu2013}
C.~W. Hsu, {\it et~al.\/}, {\it Nature\/} {\bf 499}, 188 (2013).

\bibitem{PhysRevLett.113.257401}
B.~Zhen, C.~W. Hsu, L.~Lu, A.~D. Stone, M.~Solja\ifmmode \check{c}\else \v{c}\fi{}i\ifmmode~\acute{c}\else \'{c}\fi{}, {\it Phys. Rev. Lett.\/} {\bf 113}, 257401 (2014).

\bibitem{PhysRevResearch.3.L022013}
S.~J. Palmer, V.~Giannini, {\it Phys. Rev. Res.\/} {\bf 3}, L022013 (2021).

\bibitem{PhysRevLett.131.053802}
S.~Xu, Y.~Wang, R.~Agarwal, {\it Phys. Rev. Lett.\/} {\bf 131}, 053802 (2023).

\bibitem{PhysRevB.108.085116}
S.~Vaidya, A.~Ghorashi, T.~Christensen, M.~C. Rechtsman, W.~A. Benalcazar, {\it Phys. Rev. B\/} {\bf 108}, 085116 (2023).

\bibitem{doi:10.1126/sciadv.aaw4137}
N.~Parappurath, F.~Alpeggiani, L.~Kuipers, E.~Verhagen, {\it Science Advances\/} {\bf 6}, eaaw4137 (2020).

\end{thebibliography}

\bibliographystyle{Science}


\clearpage

\end{document}